\documentclass[aps,prd,twocolumn,floatfix,nofootinbib,showpacs,superscriptaddress]{revtex4-1}

\pdfoutput=1
\usepackage{amsmath,amsfonts,amssymb,bm,ulem}

\usepackage{graphicx}
\usepackage{color}
\usepackage{url} 
\usepackage{threeparttable}
\usepackage{makecell}
\usepackage{multirow}

\definecolor{purple}{rgb}{0.1,0.5,0.4}

\usepackage[
pdfauthor={derajan},
pdfstartview=XYZ,
bookmarks=true,
colorlinks=true,
linkcolor=blue,
urlcolor=blue,
citecolor=blue,
pdftex,
bookmarks=true,
linktocpage=true,   
hyperindex=true
]{hyperref}

\begin{document}

\title{Primordial Black Holes from Domain Wall Density Fluctuations: Bridging Gravitational Wave Observations Across Two Frequency Bands}

\author{Bo-Qiang Lu} 
\email{bqlu@zjhu.edu.cn}
\affiliation{School of Science, Huzhou University, Huzhou, Zhejiang 313000, P. R. China}
\affiliation{Zhejiang Key Laboratory for Industrial Solid Waste Thermal Hydrolysis Technology and Intelligent Equipment}

\author{Cheng-Wei Chiang}
\email{chengwei@phys.ntu.edu.tw}
\affiliation{Department of Physics, National Taiwan University, Taipei 10617, Taiwan}
\affiliation{Physics Division, National Center for Theoretical Sciences, Taipei 10617, Taiwan}

\author{Tianjun Li}
\email{tli@itp.ac.cn}
\affiliation{School of Physics, Henan Normal University, Xinxiang 453007, P. R. China}
\affiliation{CAS Key Laboratory of Theoretical Physics, Institute of Theoretical Physics, Chinese Academy of Sciences, Beijing 100190, P. R. China}
\affiliation{School of Physical Sciences, University of Chinese Academy of Sciences, No.~19A Yuquan Road, Beijing 100049, P. R. China}

\begin{abstract}

We propose a novel mechanism for the formation of primordial black holes by demonstrating that the delayed production of isocurvature perturbations resulting from Poisson fluctuations within the domain wall network can lead to collapse and the formation of primordial black holes during the horizon crossing of domain walls.
Our findings establish a statistical relationship between the number of domains and the power spectrum of the perturbations. This relationship can be employed to constrain the symmetry of the model in light of the potential overabundance of primordial black holes.
Furthermore, by incorporating the effects of accretion, we demonstrate that the annihilation of the domain wall network at the QCD scale may provide a plausible common origin for gravitational wave observations across two distinct frequency bands.


\end{abstract}
\maketitle


\textbf{Introduction.} 
The pursuit of gravitational waves (GWs) has spanned over a century since Einstein first proposed their existence in 1916, making the recent observations of GWs in two different frequency bands a significant achievement. 
In the low-frequency band $\sim 10^{-9}$~Hz, the pulsar timing arrays (PTAs) including NANOGrav~\cite{NANOGrav:2023gor,NANOGrav:2023hvm}, EPTA~\cite{EPTA:2023fyk}, 
PPTA~\cite{Reardon:2023gzh}, and CPTA~\cite{Xu:2023wog} have recently presented the first convincing evidence for the Hellings-Downs angular correlation, 
which strongly supports the existence of nano-Hz stochastic gravitational wave background (SGWB).
(For a summary of new physics interpretations of the nano-Hz SGWB, see Ref.~\cite{Ellis:2023oxs}.)
In the high-frequency band $\sim 100$~Hz, GWs from mergers of black hole (BH) binaries are continuously observed by LIGO-Virgo-KAGRA (LVK) 
since 2015 and a fraction $f_{\rm PBH}\sim 10^{-3}$ of dark matter in the form of primordial black holes (PBHs) can potentially explain the observed BH 
mergers~\cite{Bird:2016dcv,Sasaki:2016jop}. 

Hawking and Carr~\cite{Hawking:1971ei,Carr:1974nx} first suggested that the self-gravity of the density perturbation originating in the early Universe could 
counteract the internal pressure gradient and collapse to form a PBH if the overdense region ceased to expand at a horizon-size scale $R_c\sim t$ 
(where $t$ denotes the cosmic time). The formation of PBHs is commonly associated with scalar inflation as the source of the density perturbations.
To account for the BH merger events by LVK, the spectral index of the power spectrum is expected to transition from a blue tilt at the QCD scale to a red tilt at the CMB scale to significantly generate the PBH abundance while remaining consistent with the CMB observations~\cite{Sasaki:2018dmp}. 
However, this requirement cannot be satisfied within the framework of single-field slow-roll inflation. It may be achievable in more complex inflationary scenarios, such as multi-scalar inflation models (see Refs.~\cite{Stewart:1997wg,Drees:2011hb,Alabidi:2012ex,Silk:1986vc,Randall:1995dj,GarciaBellido:1996mdl}).
Other sources for PBH production include the collapse of topological defects such as cosmic strings~\cite{Hawking:1987bn,Polnarev:1988dh,Brandenberger:2021zvn} 
and closed domain walls (DWs)~\cite{Ipser:1983db,Rubin:2000dq,Ferrer:2018uiu,Liu:2019lul,Gouttenoire:2023gbn,Ge:2023rrq},
the density perturbation from Q-balls during a matter-dominated era~\cite{Cotner:2016cvr,Cotner:2019ykd},
the collision of bubble walls~\cite{Crawford:1982yz,Kodama:1982sf,Hawking:1982ga,Khlopov:1998nm,Johnson:2011wt,Kusenko:2020pcg},
the collapse of false-vacuum patches during a first-order phase transition (PT)~\cite{Baker:2021sno,Gross:2021qgx,Kawana:2021tde,Liu:2021svg,Baldes:2023rqv,Flores:2024lng,Gouttenoire:2023naa,Lewicki:2023ioy,Lewicki:2024ghw,Ai:2024cka}, and the collapse of isocurvature perturbations from cold dark matter~\cite{Passaglia:2021jla} and massless scalar~\cite{Yoo:2021fxs}
(see Refs.~\cite{Sasaki:2018dmp,Carr:2020gox} for a recent review on the PBH formation scenarios).


In this Letter, we propose a novel mechanism for the formation of PBHs that is grounded in the Poisson fluctuations of the DW 
number density. We demonstrate that, in comparison to inflation, the presence of DWs can also serve as a notable source of (super)horizon-size density 
perturbations. Additionally, our approach diverges from the conventional understanding of PBH formation through the collapse of closed DWs when their radii contract to the Schwarzschild radius~\cite{Ferrer:2018uiu}. Instead, we focus on the critical collapse of horizon-sized perturbations at the time of PBH formation.

It is widely recognized that explaining phenomena beyond the standard model, such as dark matter~\cite{Bertone:2016nfn}, the strong CP problem~\cite{Marsh:2015xka}, and neutrino mass~\cite{King:2014nza}, necessitates the incorporation of new physics. DWs are sheet-like topological defects that emerge from the spontaneous breaking of a discrete symmetry in new physics models, as per the Kibble mechanism~\cite{Kibble:1976sj}.
In the accompanying paper~\cite{Lu:2024ngi}, we demonstrate that our proposed scenario enables the establishment of limits on the number of degenerate vacua through the consideration of the overabundance of PBHs, thereby providing a framework for constraining new physics models based on PBH observations. Furthermore, we demonstrate for the first time that both the nano-Hz SGWB and black hole merger events may potentially originate from the annihilation of the DW network during the QCD PT.


\textbf{PBH formation criterion.}
In the conventional framework for PBH formation, it is generally thought that adiabatic perturbations arising from inflation play a crucial role in generating density perturbations. At an initial time $t_i$, the superhorizon mode of these perturbations may have the potential to form a PBH at a later time $t_f$, particularly during the horizon crossing phase, provided that the density fluctuations induced by the adiabatic perturbation surpass a critical threshold. Additionally, it is important to note that the power spectrum for the superhorizon mode needs to be sufficiently robust to effectively lead to the formation of PBHs.
In this Letter, we shift our focus from adiabatic perturbations to the isocurvature perturbations originating from the Poisson fluctuations within the DW network. It is noteworthy that while the superhorizon mode of the isocurvature perturbation remains constant in a radiation-dominated Universe, the induced density perturbation experiences growth over time. Furthermore, we highlight that isocurvature perturbations are not generated immediately at the time of DW formation; instead, their production is delayed until the moment when the closed DWs enter the horizon.


Let's consider a two-component Universe comprising radiation and DWs, with the assumption that the radiation is the dominant component and 
the perturbations are primarily induced by the DW network. 
In contrast to the commonly employed co-moving curvature perturbation $\mathcal{R}$~\cite{Green:2004wb,Shandera:2012ke,Young:2014ana},
we adopt the traditional density contrast $\delta=\delta\rho_{\rm tot}/\bar{\rho}_{\rm tot}$ to estimate the PBH formation,
where $\delta\rho_{\rm tot}=\delta\rho_w+\delta\rho_{r}\simeq \delta\rho_w$ (with $\delta\rho_i=\rho_i-\bar{\rho}_i$)
and $\bar{\rho}_{\rm tot}=\bar{\rho}_{w}+\bar{\rho}_{r}\simeq \bar{\rho}_{r}$ are the total density perturbation and background energy density, respectively,
and the subscripts $w$ and $r$ stand for wall and radiation, respectively. 
The criterion for the formation of PBHs during the horizon crossing, as explained by Hawking and Carr in Ref.~\cite{Hawking:1971ei,Carr:1974nx}, 
can be clearly represented by the Newtonian potential
\begin{equation}\label{eq:phi}
    \Phi\simeq -\frac{3}{2}\left(\frac{\mathcal{H}(t_f)}{k}\right)^2\delta(t_f),
\end{equation}
where the Newton gauge is assumed.
During the horizon crossing $k\simeq \mathcal{H}(t_f)$ at the time $t_f$, the overdense region could collapse to form a PBH if $|\Phi|\gtrsim 1$, 
which corresponds to the requirement that $\delta\gtrsim \delta_c\simeq 0.5$.

We can introduce the entropy perturbation $S(t)=\delta_w/(1+\omega_w)-\delta_{r}/(1+\omega_r)$, where $\delta_w=\delta\rho_w/\bar{\rho}_w$, $\delta_r=\delta\rho_r/\bar{\rho}_r$, 
and $\omega_w$ and $\omega_r$ are the equation of states (EOS) of DW and radiation, respectively. 
In the absence of anisotropic stress, the evolution of $S$ is described by the Kodama-Sasaki equation~\cite{Lu:2024ngi,Kodama:1986ud}
\begin{equation}\label{eq:KSE}
    \frac{S^{\prime \prime}}{\mathcal{H}^{2}}+\frac{(1-3\mu_2)S^{\prime}}{\mathcal{H}}=-\left(\frac{k}{\mathcal{H}}\right)^2\left(\mu_1\delta^C+\mu_2 S\right),
\end{equation}
where the prime denotes the derivative concerning the conformal time, $\mu_1$ and $\mu_2$ are determined by $\omega_r$, $\omega_w$ and $f_w$ (see~\cite{Lu:2024ngi}), $\omega_{\rm eff}=\bar{p}_{\rm tot}/\bar{\rho}_{\rm tot}$, $c_s^2=\bar{p}_{\rm tot}^{\prime}/\bar{\rho}_{\rm tot}^{\prime}$, respectively, and $\delta^C=\delta+3\left(\frac{\mathcal{H}}{k}\right)(1+\omega_{\rm eff}) v$, with $v$ being the fluid velocity. 
Considering the superhorizon mode with $k\ll \mathcal{H}(t_i)$ at the initial time $t_i$ in a radiation-dominated Universe, 
we can neglect the term on the right-hand side of Eq.~\eqref{eq:KSE}. Consequently, the solution becomes constant, indicating that the superhorizon 
isocurvature perturbation remains unchanged, represented by $S\equiv S(0)$, prior to the horizon crossing.
For the case where $\delta_{r}\simeq 0$, the isocurvature perturbation is expressed as $S(t)\simeq 3\delta_w(t)$. 
The superhorizon mode of the density contrast is then given by $\delta(t)\simeq f_w(t)S(0)/3$, 
where $f_w(t)=\bar{\rho}_w/\bar{\rho}_{r}=32\pi G\mathcal{A}\sigma_w t/3$ represents the energy fraction of the wall~\cite{Lu:2024ngi}.
For an initial isocurvature perturbation $S(0)\sim \mathcal{O}(1)$, we find that the condition $\delta\gtrsim \delta_c$ can be satisfied when DWs
annihilate at a sufficiently late time, allowing the energy fraction of the DWs to grow to approximately $f_w(t)\sim \mathcal{O}(0.1)$. 


We will demonstrate that the production of isocurvature perturbations from the DW network can be delayed until the DWs re-enter the horizon, allowing for a significant $f_w(t)$. Additionally, we will discuss how the initial condition $S(0)\sim \mathcal{O}(1)$ can arise from Poisson fluctuations in the DW network.

\textbf{Delayed isocurvature perturbation from Poisson fluctuation.}
The DWs are established as a radiation gas with $\omega_w=1/3$ at the initial time. Following their formation, the DW expands rapidly to a superhorizon size while remaining ultra-relativistic due to curvature acceleration resulting from the surface tension of the wall \cite{Martins:2016ois,Lu:2024ngi}. Throughout this period, the behavior of the DWs resembles that of a radiation species; consequently, no isocurvature perturbations are produced.
As the correlation length, or curvature radius, of the DW expands to a superhorizon scale, the effects of chopping between walls and Hubble friction due to cosmic expansion become increasingly pronounced. The presence of Hubble friction may induce deceleration of the wall, while the chopping effect can result in a reduction of wall size and the formation of closed configurations. Such closed configurations are predominantly established during the wall's evolution \cite{Lazanu:2015fua}, which can significantly attenuate the chopping effect. As a result, the size of the closed DW consistently remains at a superhorizon scale and continues to increase until the occurrence of DW annihilation, as evidenced by recent literature \cite{Deng:2020mds,Gouttenoire:2023gbn}.
Isocurvature perturbations can only arise when the correlation length of the DWs is reduced to the horizon scale. This limitation occurs because DWs that are separated by a superhorizon correlation length are causally disconnected, resulting in a lack of perturbations within a horizon volume. The introduction of a bias potential, $V_{\rm bias}$, which disrupts the discrete symmetry, can trigger the annihilation of the DW network. Under the influence of this potential pressure, the size of the closed DWs tends to decrease and re-enter the horizon as $V_{\rm bias}$ becomes dominant. Therefore, the generation of isocurvature perturbations due to DW fluctuations is effectively postponed until the inception of the wall's annihilation.

The DWs tend to increase their size and merge nearby walls, giving rise to a network of DWs characterized by larger sizes and a reduced wall number at the 
later stages of evolution. 
For the spontaneous breakdown of a $Z_N$ symmetry, simulations~\cite{Press:1989yh,Hiramatsu:2012sc,Lazanu:2015fua} have indicated 
that there eventually exist about $N$ horizon-size DWs that separate distinguishable domains.
The statistics of a few-body system can be accurately represented by the Poisson distribution. 
Consider a network consisting of $N$ identical walls in a Hubble volume 
$V_H\equiv 4\pi r_H^3/3$ (where $r_H=1/H$ is the Hubble horizon).  Then the probability of finding $n$ walls in a volume of $V\equiv 4\pi r^3/3$ is given by 
\begin{equation}\label{eq:PD}
    \mathbb{P}_n(V)=\left(\frac{V}{V_w}\right)^{n}\frac{e^{-V/V_w}}{n!},
\end{equation}
where $V_w=4\pi L_w^3/3$ is the volume of the wall and $V=nV_w$, with $L_w$ being the DW correlation length.
The mean square energy density fluctuations relative to the total energy density in the volume $V$ can be determined due to Poisson fluctuations of the DWs as
\begin{equation}
\label{eq:MSdelta}
    \langle \delta^2(r) \rangle=\frac{\langle (\delta \bar{\rho}_{w}(r))^2 \rangle}{\rho_{\rm tot}^2}=\frac{\langle \bar{\rho}_w\rangle^2V_w}{\rho_{\rm tot}^2V}.
\end{equation}
We further consider a Gaussian distribution for $\delta$
\begin{equation}
\label{eq:gaussP}
    P(\delta)=\frac{1}{\sqrt{2\pi}\sigma}\exp\left(-\frac{\delta^2}{2\sigma^2}\right),
\end{equation}
where the variance of the probability distribution is determined by Eq.~\eqref{eq:MSdelta}. 
In a radiation-dominated Universe with $f_w \ll 1 $ and DWs with a horizon-sized correlation length $L_w(t_e)\sim t_e\sim 1/H(t_e)$, where $t_e$ denotes the time when the DW enters into the horizon, the variance of the perturbation is given by
\begin{eqnarray}\label{eq:vardw}
    \sigma^2 = \left. \langle \delta^2 \rangle \right|_{V \to V_H} = \frac{f_w^2}{N}.
\end{eqnarray}

For a Poisson distribution, the power spectrum of the isocurvature perturbation generated by correlations of local fluctuations from horizon-sized DWs is given by \cite{Lu:2024ngi}
\begin{equation}
    \mathcal{P}_{S}(k)=\frac{6}{\pi}\left(\frac{k}{k_e}\right)^3\simeq \left(\frac{k}{\mathcal{H}(t_e)}\right)^3~~{\rm for~~}k<k_e,
\end{equation}
where $k_e=a/L_w(t_e)\sim \mathcal{H}(t_e)$ and $t_e$ denotes the horizon crossing time of the wall. 
Hence, the power spectrum of the density perturbation can be estimated as $\sigma(k)^2 = \mathcal{P}_{\delta}(k)\simeq f_w(t_{f})^2(k/\mathcal{H}(t_{e}))^3$.
We observe that the variance in Eq.~\eqref{eq:vardw} can replicate the power spectrum of the horizon mode with $k\simeq \mathcal{H}(t_e)$, differing only by a factor of $1/\sqrt{N}$ to account for the ensemble average of the fluctuation.
Utilizing Eq.~\eqref{eq:phi} and recognizing that $\Psi=\Phi$ (where $\Psi$ denotes the curvature perturbation) in a radiation-dominated Universe, the power spectrum of $\Psi$ during the horizon crossing of mode $k$ can be expressed as
\begin{equation}\label{eq:csp}
    \mathcal{P}_{\rm \Psi}(k)\simeq f_w(t_f)^2\left(\frac{k}{\mathcal{H}(t_e)}\right)^3\left(\frac{\mathcal{H}(t_f)}{k}\right)^4
    ~~{\rm for}~~k\gtrsim \mathcal{H}(t_f).
\end{equation}

Using Eqs.~\eqref{eq:gaussP} and~\eqref{eq:vardw}, we have $\bar{\delta}=\int_0^{\infty}\delta P(\delta)d\delta=f_w/\sqrt{2\pi N}$ and $\bar{S}(0)=3/\sqrt{2\pi N}$. This analysis demonstrates that the initial averaged isocurvature perturbation $\bar{S}(0)\sim \mathcal{O}(1)$ for superhorizon modes can be effectively realized within a network comprising $N\lesssim 10$ DWs.
As illustrated, when the pressure resulting from the bias potential becomes comparable to the surface tension, the superhorizon-sized DWs rapidly contract to horizon size and subsequently annihilate. Therefore, it can be stated that $t_{e}\sim t_{\rm ann}$, where $t_{\rm ann}\simeq \sigma_w/\Delta V_{\rm bias}$ denotes the DW annihilation time.
From the curvature power spectrum described in Eq.~\eqref{eq:csp}, it can be observed that PBHs can be produced significantly at the horizon crossing time of the DWs, specifically at \(t_f = t_e \sim t_{\text{ann}}\), provided that the energy fraction of the DWs is approximately \(f_w(t_e) \simeq \mathcal{O}(0.1)\) when the DWs enter the horizon. Additionally, the condition \(|\Phi| \gtrsim 1\), which is necessary for the PBH formation, can also be satisfied.
Note that the Newtonian potential criterion for PBH formation can be applied to isocurvature perturbations as the potential increases over time. The initial isocurvature perturbation becomes adiabatic when the over-density exceeds the critical value \(\delta_c\), causing the overdense region to stop expanding and become gravitationally self-bound.

\textbf{PBH formation and evolution.}
The PBH mass from critical collapse follows a simple scaling law near the threshold $M=CM_H(\delta-\delta_c)^{\gamma}$~\cite{Choptuik:1992jv,Evans:1994pj,Koike:1995jm}, 
where we take $C=3.3$ and $\gamma=0.36$ and note that these parameters may be affected by interior pressure gradients of matter~\cite{Musco:2023dak}. 
Ref.~\cite{Yoo:2021fxs} shows that the isocurvature and adiabatic perturbations could lead to similar behaviors in the evolution of the PBH mass near the critical threshold
in a radiation-dominated Universe.
The total mass contained in a Hubble horizon volume is $M_H=1/(2GH)$. Using the Press-Schechter formalism, the PBH energy 
fraction $\beta\equiv \rho_{\rm PBH}(t_f)/\rho_{\rm tot}(t_f)$ at the formation time $t_f$ can be estimated as~\cite{Byrnes:2018clq}
\begin{equation}\label{eq:bt}
    \beta(M)=2\int_{\delta_c}^{\infty}C(\delta-\delta_c)^{\gamma}P(\delta)d \delta.
\end{equation}
The PBH mass function is then given by~\cite{Byrnes:2018clq,Niemeyer:1997mt}
\begin{equation}\label{eq:MF}
    \psi(M)=\frac{2}{\Omega_{\mathrm{CDM}}} \int_{-\infty}^{\infty} P(\delta) 
    \frac{MB^{1 / \gamma}}{\gamma M_{H}}  \sqrt{\frac{M_{\mathrm{eq}}}{M_{H}}} d \ln M_{H},
\end{equation}
where $B=M/(CM_H)$, $\Omega_{\rm CDM}=0.25$ is the dark matter relic abundance, and $M_{\rm eq}\simeq 2.8\times 10^{17}M_{\odot}$ is the horizon mass at the time of matter-radiation equality.
The PBH relic abundance is related to $\psi(M)$ by 
\begin{equation}
    \Omega_{\mathrm{PBH}}=\Omega_{\mathrm{CDM}} \int \psi(M) d \ln M
\end{equation}
and the average PBH mass is defined as 
$\langle M\rangle=\Omega_{\mathrm{CDM}} \int M\psi(M) d \ln M/\Omega_{\rm PBH}$.
The critical value $\delta_c$ is contingent on the EOS of the Universe, see Fig.~3 in Ref.~\cite{Lu:2024ngi}.
Fitting to the simulations~\cite{Musco:2012au}, we find the formula 
\begin{equation}\label{eq:fit}
    \delta_c=\frac{\delta_0}{\left[ 1+(\omega_0/\omega_{\rm eff})^{s_1} \right]^{s_2}},
\end{equation}
where the fit parameters $\{\delta_0,\omega_0,s_1,s_2 \}$ are given in Table~I of Ref.~\cite{Lu:2024ngi}. We also take into account the variation of the EOS during the PT~\cite{Byrnes:2018clq} and adopt the simulation results with a Gaussian perturbation spatial profile~\cite{Musco:2012au}.

In our scenario, non-spherical effects on the PBH formation are determined to be negligible. First, simulations reveal that the spatial distribution of DWs is isotropic, with closed structures predominantly forming during the DW evolution~\cite{Lazanu:2015fua}. Second, further simulations~\cite{Yoo:2020lmg,Escriva:2024aeo} indicate that non-spherical effects play an insignificant role in PBH formation in the radiation-dominated Universe, as the collapse process is primarily governed by gas pressure~\cite{Harada:2013epa}.

\begin{figure}
    \centering
    \includegraphics[width=85mm,angle=0]{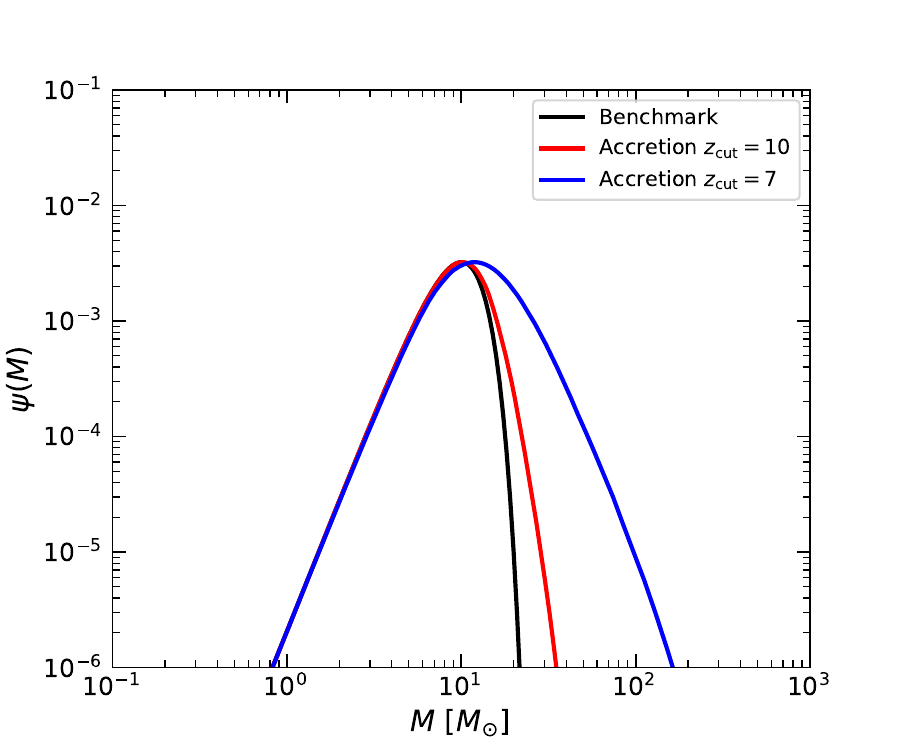}
    \caption{PBH mass function, assuming $(T_{\rm ann}, f_w(T_{\rm ann})) = [0.1~{\rm GeV}, 0.1]$ and the DW number $N=3$.}
    \label{fig:fraction}
\end{figure}

PBHs can accrete the surrounding mass and grow over time after their formation. 
The PBH accretion becomes significant during the CMB epoch at redshift around $10\lesssim z\lesssim 30$.
We refer to Appendix~\ref{app:PBHaccreation} for more details on the PBH accretion. 
In Fig.~\ref{fig:fraction}, we present the mass function of PBHs resulting from fluctuations in DW number density at the benchmark point 
of $(T_{\rm ann}, f_w(T_{\rm ann})) = [0.1~{\rm GeV}, 0.1]$ with $N=3$ 
(see the Appendix~\ref{app:models} for a particle physics model implementation).
The black curve illustrates the PBH mass spectrum without considering 
the accretion effect, while the red and blue curves show the results that account for PBH accretion with redshift cut-offs 
$z_{\rm cut}=10$ and $z_{\rm cut}=7$, respectively, below which the accretion becomes negligible.

\begin{figure}
    \centering
    \includegraphics[width=85mm,angle=0]{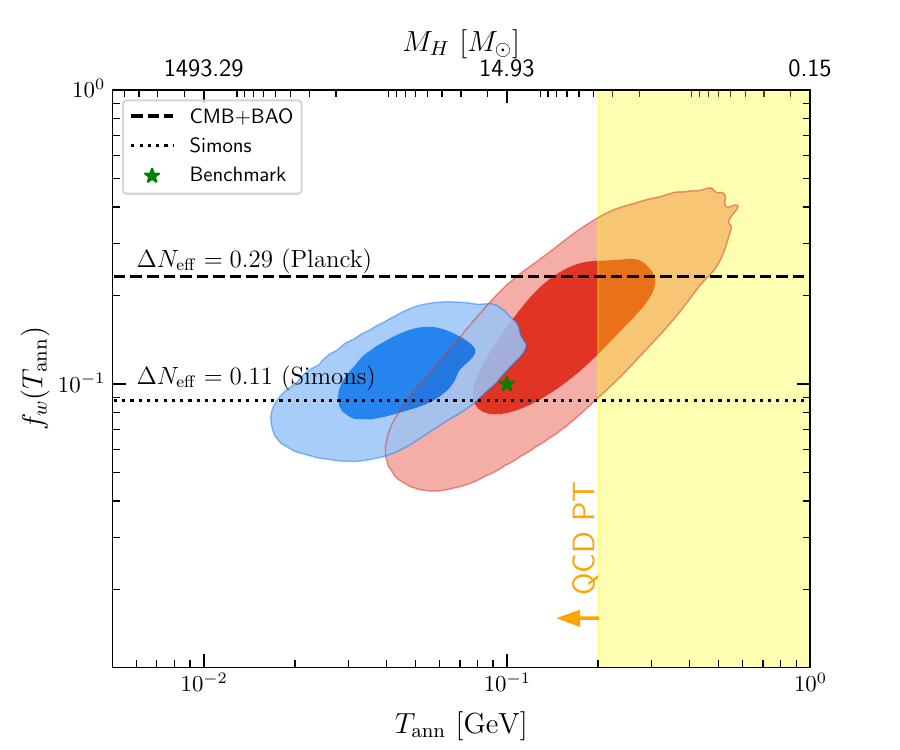}
    \caption{The red and blue $1\sigma$ and $2\sigma$ regions represent the parameter space of a fit to the NANOGrav 15-year dataset~\cite{NANOGrav:2023gor} 
    and IPTA-DR2 dataset~\cite{Antoniadis:2022pcn} with the DW annihilation scenario, respectively.
    Depending on the model, DWs may emit relativistic dark radiation and contribute to the effective number of neutrino species 
    with $\Delta N_{\rm eff}\simeq f_w(T_{\rm ann})(p_{\rm DR}/0.16)(10.75/g_{*,s})^{1/3}$ (where $p_{\rm DR}=0.16$ is adopted). 
    The constraints on $\Delta N_{\rm eff}$ from current Planck~\cite{Planck:2018vyg} 
    and future Simons~\cite{SimonsObservatory:2018koc} experiments are represented by the dashed and dotted lines, respectively.}
    \label{fig:N3}
\end{figure}

\textbf{GW phenomenology.}
LVK has documented over one hundred BH merger events. While favoring BHs of astrophysical origin, some events may not align with stellar evolution predictions. 
For instance, GW190425~\cite{LIGOScientific:2020aai}, GW190814~\cite{LIGOScientific:2020zkf}, and GW230529\_181500~\cite{LIGOScientific:2024elc} may have component 
masses exceeding neutron star or stellar BH upper/lower limits~\cite{Clesse:2020ghq,Chen:2021nxo}, while the component masses of GW190521 fall within the mass 
gap $\sim (50-135)~M_{\odot}$ of the stellar-origin BHs~\cite{Palmese:2020xmk}, implying a primordial origin for this event~\cite{DeLuca:2020sae}. 
Using the PBH mass spectrum~\eqref{eq:MF} we find that $(\langle M\rangle,~f_{\rm PBH})\sim [(10-15)M_{\odot},~(2-4)\times 10^{-3}]$ 
at the benchmark point (the green star in Fig.~\ref{fig:N3}), depending on the choices of $z_{\rm cut}$ for the PBH accretion. 
As a reference, the Bayesian analysis of LVK O3 events in Ref.~\cite{Andres-Carcasona:2024wqk} favors the parameters $[16.8M_{\odot},~1.1\times 10^{-3}]$ for the PBH origin.
It is expected that the DW should annihilate before they dominate the Universe. The accompanying GWs have been considered as a potential source for the recent PTA 
observations~\cite{Chiang:2020aui,Lu:2023mcz,Ferreira:2022zzo,Blasi:2023sej}.
We follow the model-independent procedure outlined in the joint paper~\cite{Lu:2024ngi} to fit the PTA datasets and the main results are depicted in Fig.~\ref{fig:N3}. 
We observe that the DW annihilation interpretation for the PTA observations requires that the annihilation of DWs occurs around the QCD scale.
The QCD instanton effect due to the anomaly of the discrete symmetry under the QCD gauge group can naturally lead to the annihilation of DW 
at the QCD scale~\cite{Preskill:1991kd,Lu:2023mcz}.
Therefore, our findings suggest the DW as a common origin for both the nano-Hz SGWB in PTAs and BH mergers events by LVK.


It is worth mentioning that the PBH mass distribution~\eqref{eq:MF} has a cut-off at high masses due to the DW annihilation, 
which could significantly enhance the Bayesian evidence for LVK ovservations~\cite{Andres-Carcasona:2024wqk,Gow:2019pok,Hall:2020daa}. 
We also note that the constraints on new physics interpretations for PTAs from the over-production of PBHs with masses range from about $10^{-3}~M_{\odot}$ to $1~M_{\odot}$ 
have been discussed in previous literatures, including the collapses of the scalar perturbations~\cite{Chen:2019xse,Franciolini:2023pbf}, 
the closed DW~\cite{Ipser:1983db,Ferrer:2018uiu,Gouttenoire:2023ftk}, and false-vacuum patches during a first-order 
PT~\cite{Gross:2021qgx,Liu:2021svg,Lewicki:2023ioy,Flores:2024lng,Gouttenoire:2023bqy}.

\textbf{Applications to new physics models.}
We have identified a significant relationship between the number of domains in the model and the variance of the density contrast as described by 
Eq.~\eqref{eq:vardw}. Consequently, our scenario can be directly utilized to constrain the model's symmetry, which is a fundamental aspect of model formulation, by using observations of the PBH relic abundance.
As an example, we demonstrate in Ref.~\cite{Lu:2024ngi} that the DW annihilation in a model with $Z_2$ symmetry, commonly discussed in the literature, is ruled out as an explanation for the PTA observations due to the overabundance of PBHs. Additionally, $Z_N$ symmetric models with $N\lesssim 10$ face strong constraints from PTA observations.

\textbf{Conclusions.}
In this Letter, we have proposed that the number density fluctuations within a DW network can serve as a novel source of density perturbations. The generation of isocurvature perturbations originating from the Poisson fluctuations of the DW network is delayed until the horizon crossing of the closed DWs. This mechanism can ultimately result in critical collapse and subsequent formation of PBHs. 
By considering a benchmark model, we demonstrated that the mass distribution of PBHs resulting from DW Poisson fluctuations can 
explain the BH merger events observed by LVK, while the annihilation of DWs can naturally generate the nano-Hz SGWB in PTAs. 
Future GW observations from LVK, space-based interferometer, and PTAs~\cite{DeLuca:2020sae,Chiang:2019oms,Chiang:2020aui} as well as collider 
experiments~\cite{Baldenegro:2018hng,Bauer:2018uxu,Ferreira:2022zzo} could further test our scenario.


\textbf{Acknowledgments.}
The authors appreciates Misao Sasaki for helpful comments and suggestions.
BQL is supported by the National Natural Science Foundation of China (NSFC) under Grant No. 12405058 and by the Zhejiang Provincial Natural Science Foundation of China under Grant No.~LQ23A050002. 
CWC is supported by the National Science and Technology Council under Grant No.~NSTC-111-2112-M-002-018-MY3. 
TL is supported by the National Key Research and Development Program of China Grant No. 2020YFC2201504, by Project No.~12275333 supported by the NSFC, by the Scientific Instrument Developing Project of the Chinese Academy of Sciences, Grant No.~YJKYYQ20190049, and by the International Partnership Program of the Chinese Academy of Sciences for Grand Challenges, Grant No.~112311KYSB20210012.

\appendix
\twocolumngrid
\section{PBH accretion}\label{app:PBHaccreation}
In this appendix, we present the details regarding PBH accretion and the model implementation used in our study.

After their formation, PBHs can accrete surrounding mass and grow over time while also losing mass through the Hawking radiation. 
According to the generalized second law of thermodynamics, the accretion of PBHs would dominate the radiation era when the thermal bath temperature exceeds that of the BHs, 
i.e., $T\gtrsim T_{\rm BH}$, where $T_{\rm BH}=1/(8\pi GM)$ represents the BH temperature. Zel'dovich and Novikov suggested that horizon-size PBHs could grow at 
a rate similar to the cosmological horizon before the end of the radiation era based on the Bondi-type accretion~\cite{Zeldovich:1967lct}. 
However, subsequent research in Ref.~\cite{Carr:1974nx} did not find a self-similar solution connecting a BH to an exact flat Friedmann background via a sound wave in a radiation-dominated Universe. The self-similar growth of PBHs may lead to a population with a mass around $10^{17}M_{\odot}$, but this has not been confirmed by observations~\cite{Carr:2010wk}. Therefore, it is widely believed that horizon-size PBHs do not grow in the radiation era~\cite{Carr:2020gox}. 

The spherical accretion of an isolated PBH surrounded by a dark matter halo and/or a hydrogen gas can be described by the Bondi-Hoyle rate 
\begin{equation}
    \dot{M}=\lambda 4 \pi r_{\mathrm{B}}^2 \rho_{\mathrm{gas}} v_{\mathrm{eff}},
\end{equation}
where $\rho_{\rm gas}$ is the density of the hydrogen gas, $v_{\rm eff}=\sqrt{v_{\rm rel}^2+c_{s}^2}$ is the effective velocity, $c_s$ is the sound speed, and $r_{\rm B}\equiv M/v_{\rm eff}^2$ is the Bondi-Hoyle radius. The accretion parameter $\lambda$ takes into account the effects of gas viscosity, the growth of a dark halo, the Hubble expansion, and the Compton scattering between the CMB radiation and the hydrogen gas. We refer to 
Refs.~\cite{Ricotti:2007au,Ricotti:2007jk,DeLuca:2020fpg,DeLuca:2020qqa} for more details. The accretion rate can be estimated by using~\cite{Ricotti:2007au}
\begin{equation}
    \dot{m}=0.023 \lambda\left(\frac{1+z}{1000}\right)\left(\frac{M}{M_{\odot}}\right)\left(\frac{v_{\text {eff }}}{5.74 \mathrm{~km} \mathrm{~s}^{-1}}\right)^{-3}.
\end{equation}
For redshift $z\lesssim z_{\rm dec}\simeq 130$, where the baryonic matter decouples from the CMB radiation, $\dot{m}$ remarkably increases due to the decrease of the relative velocity $v_{\rm rel}$ between PBHs and gas. Depending on the PBH mass, $\dot{m}$ peaks at redshift $z\sim 30$, with an accretion time scale of $t_{\rm Salp}\sim 10^8$~yr, much larger than the age of the Universe $t_{\rm age}\sim 10^7$~yr. Therefore, the accretion of the PBH inside a virialized halo is significant during the redshift $10\lesssim z\lesssim 30$. For $z\lesssim 10$, the structure formation begins and the relative velocity increases to $v_{\rm rel}\gtrsim 10$~km/s, leading to a decrease of PBH accretion~\cite{Ricotti:2007au}. Other factors, such as reionization and global feedback, can also decrease the PBH accretion rate~\cite{Ricotti:2007au,DeLuca:2020fpg}. To address modeling uncertainties, we consider cut-off redshift values 
of $z_{\rm cut}=10$ and 7, as adopted in Refs.~\cite{DeLuca:2020fpg} and~\cite{Ricotti:2007au}, respectively, below which the accretion is negligible.

In Table~\ref{tab:iii}, we show the average mass $\langle M\rangle$ and the PBH density fraction $f_{\rm PBH}$ around the benchmark point $(f_w(T_{\rm ann},T_{\rm ann})=(0.1,100{\rm MeV})$, assuming $N=3$. We observe that the accretion can be significant for PBH mass $M\gtrsim 10~M_{\odot}$. Therefore, a smaller $z_{\rm cut}$ could lead to a larger accretion effect, leaving significant impacts on the interpretation of the LIGO observations.

\begin{table}[tbp]
    \centering
    \renewcommand\arraystretch{1.5}
    \begin{tabular}{cccc}
    \hline
    \multirow{2}*{~~$\left(T_{\rm ann}/{\rm MeV},f_w(T_{\rm ann})\right)$~~} & 
    \multicolumn{3}{c}{$(\langle M\rangle/M_{\odot},~f_{\rm PBH}/10^{-3})$} \\
    \cline{2-4}
     & ~~no acc.~~ & ~~$z_{\rm cut}=10$~~ & ~~$z_{\rm cut}=7$~~ \\
    \hline 
    (100,~0.1) & (9.16,~2.6) & (9.79,~2.9) & (14.1,~4.1) \\
    (100,~0.098) & (9.03,~0.9) & (9.62~1.0) & (13.6,~1.4) \\
    (100,~0.102) & (9.28,~6.9) & (9.97,~7.6) & (14.6,~10.9) \\
    (80,~0.1) & (14.3,~1.8) & (20.9,~2.9) & (82.4,~5.9) \\
    (120,~0.1) & (6.37,~4.9) & (6.45,~4.9) & (7.26,~5.7) \\
    \hline
    \end{tabular}
    \caption{\label{tab:iii} The average mass $\langle M\rangle$ and the PBH density fraction $f_{\rm PBH}$ around the benchmark point $(f_w(T_{\rm ann},T_{\rm ann})=(0.1,100{\rm MeV})$, assuming $N=3$.}
  \end{table}

\section{Model implementation}\label{app:models}

Let's now construct a concrete particle physics model whose domain walls are annihilated by the QCD instanton-induced bias potential. Such a scenario could naturally generate the nano-Hz gravitational waves by the domain wall annihilation. Furthermore, the density perturbations from the Poisson fluctuations in the domain wall network could collapse to form PBHs that may account for the LIGO BH-BH merger events.

Consider a scalar potential~\cite{Chiang:2019oms}
\begin{equation}\label{eq:potenphi}
    V(\Phi)=\lambda\left( \Phi^{\dagger}\Phi-\frac{v_{\phi}^2}{2} \right)^2-\frac{\mu_3}{\sqrt{2}}\left( \Phi^3+\Phi^{\dagger 3} \right),
\end{equation}
where the scalar field $\Phi$ is decomposed as  
\[
\Phi = \frac{1}{\sqrt{2}}(\phi + v_{\phi})e^{ia/v_{\phi}},
\]  
where $\phi$ and $a$ represent the radial and angular (axion) components, respectively, and $v_{\phi}$ is the vacuum expectation value (VEV) of $\Phi$. The parameter $\mu_3$ is assumed to be real.  
In Eq.~\eqref{eq:potenphi}, the first term on the right-hand side respects a global $U(1)$ symmetry, which is softly broken to a $Z_3$ symmetry by the second term. 

To realize our framework, we consider the Kim-Shifman-Vainshtein-Zakharov (KSVZ) axion model~\cite{Kim:1979if,Shifman:1979if}. This model introduces a heavy vector-like quark $Q$ and a scalar field $\Phi$. Under the standard model gauge group $SU(3)_c \times SU(2)_L \times U(1)_Y$, the charges of $Q$ and $\Phi$ are $(3,1,0)$ and $(1,1,0)$, respectively. The transformations of $Q$ and $\Phi$ under the global $Z_3$ symmetry are given by
\begin{equation}
    \Phi\to e^{i\alpha}\Phi,~~
    Q_L\to e^{i\alpha/2}Q_L,~~{\rm and}~~
    Q_R\to e^{-i\alpha/2}Q_R,
\end{equation}
where $\alpha=2\pi j/\mathcal{N}$ with $j=0,1,2,...,\mathcal{N}-1$ and $\mathcal{N}=3$ for the $Z_3$ symmetry.
The spontaneous breakdown of $Z_3$ symmetry generates the potential of axion $a$:
\begin{equation}\label{eq:Va}
    V_a(a)=-\frac{1}{4}\mu_3v_{\phi}^3\cos\left( \frac{3a}{v_{\phi}}\right).
\end{equation}
This potential exhibits three-fold degenerate vacua located at $a_j/v_{\phi}=2\pi j/3$ for $j=0,~1,~2$. DWs with a surface tension of $\sigma_w\simeq 8m_af_a^2$ (where $m_a^2=9\mu_3v_{\phi}/4$ and $f_a=v_{\phi}/3$) are formed via the Kibble mechanism~\cite{Kibble:1976sj} to separate the degenerate vacua.

The Yukawa interaction for the heavy quark \( Q \) with the scalar field \( \Phi \) is expressed in the following Lagrangian:
\[
\mathcal{L}_Y=-(y_{Q}\Phi \bar{Q}_L Q_R+\text{h.c.}),
\]
where \( y_Q \) is the Yukawa coupling, and \( Q_L \) and \( Q_R \) denote the left-handed and right-handed components of the heavy quark, respectively. When the scalar field \( \Phi \) acquires a VEV \( v_\phi \), the \( Z_3 \) symmetry is spontaneously broken, leading to a mass term for the quark:
\[
\mathcal{L}_m=-m_Q \bar{Q}_L Q_R e^{ia/v_{\phi}} + \text{h.c.},
\]
with \( m_Q = \frac{y_Q v_{\phi}}{\sqrt{2}} \), and \( a \) representing the axion field.
To eliminate the phase \( a \) from this mass term, one can implement a chiral transformation:
\[
Q_L \to e^{ia/2v_\phi} Q_L, \quad Q_R \to e^{-ia/2v_\phi} Q_R.
\]
However, this transformation is anomalous under the QCD gauge group, resulting in a topological interaction between the axion and gluons, described by:
\[
\mathcal{L} \supset \frac{g_s^2 a}{32 \pi^2 v_\phi} G^{\mu\nu} \widetilde{G}_{\mu\nu},
\]
where \( G_{\mu\nu} \) is the gluon field strength tensor and \( g_s \) indicates the strong coupling constant.
After the QCD phase transition, this topological term generates an effective potential for the axion \( a \) that is capable of breaking the \( Z_3 \) symmetry via QCD instanton effects:
\begin{equation}\label{eq:Vca}
V_c(a)=-\Lambda_c^4 \cos\left(\frac{N_c a}{3f_a}+\vartheta\right),
\end{equation}
where \( \Lambda_c = (m_u \Lambda_{\text{QCD}}^3)^{1/4} \simeq 66.2 \)~MeV, and \( \vartheta \) represents the strong CP-violating phase. The parameter \( N_c \) relates to the number of quarks charged under the \( Z_3 \) symmetry and, to avoid complications linked to color anomaly degeneracies, we set \( N_c = 1 \).
The potential \( V_c(a) \) serves to lift the three-fold vacuum degeneracy by introducing a bias energy:
\begin{equation}
    V_{\text{bias}}^{1/4} = (1.5 \Lambda_c^4)^{1/4} \simeq 73.3 \text{ MeV}.
\end{equation}
This bias plays a crucial role in selecting a specific vacuum and significantly influences the dynamics of the axion field and its interactions with QCD. 

Using the relations derived in Ref.~\cite{Lu:2024ngi}
\begin{equation}\label{eq:sgfw}
    \left(\frac{\sigma_w^{1/3}}{10^5~{\rm GeV}}\right)^3=39.6f_w(T)\left(\frac{1}{\mathcal{A}}\right)\left(\frac{g_{*}(T)}{10.75}\right)^{1/2} 
    \left(\frac{T}{0.1~{\rm GeV}}\right)^{2}
\end{equation}
and
\begin{equation}\label{eq:Vbft}
    \frac{V_{\rm bias}^{1/4}}{0.1~{\rm GeV}}=1.37(f_w(T_{\rm ann}))^{1/4}\left( \frac{T_{\rm ann}}{0.1~{\rm GeV}} \right)
    \left( \frac{g_*(T_{\rm ann})}{10.75} \right)^{1/4},
\end{equation}
we obtain $\sigma_w^{1/3}\simeq 1.5\times 10^5$~GeV and $V_{\rm bias}^{1/4}\simeq 77.0$~MeV for the benchmark model. Therefore, we have confirmed that the QCD instanton-induced potential~\eqref{eq:Vca} can provide correct bias energy for the decay of the degenerate vacua~\eqref{eq:Va} in the benchmark model.

The scale $v_{\phi}\sim 10^{5}$~GeV is significantly below the Planck scale, effectively ensuring the quality of $Z_3$. To verify the conservation of strong CP of the heavy axion vacuum, we examine the minimum of the potential
\begin{equation}
    \langle \theta\rangle=\frac{m_*^2\tan\vartheta}{N_c(m_a^2+m_*^2)},
\end{equation}
where $m_*^2=\Lambda_c^4N_c^2\cos\vartheta/v_{\phi}^2$ and $\theta=a/v_{\phi}$. Assuming $\cos\vartheta\sim 0.1$, $\tan\vartheta\sim 1$ and $m_a\sim v_{\phi}\sim 10^5$~GeV, we have $\langle \theta\rangle\simeq 10^{-26}$, which is far below the experimental constraint $\theta_{\rm QCD}\lesssim 10^{-10}$~\cite{Baker:2006ts}.


\twocolumngrid
\bibliographystyle{apsrev4-1}
\bibliography{reference}

\end{document}